\documentclass{article}
\usepackage{graphics} 
\usepackage{amssymb, amsmath, times, lineno}
\usepackage{icrctc07}
\title{Systematic study 
of atmosphere-induced influences and
uncertainties on shower reconstruction at the Pierre Auger
Observatory}
\shorttitle{Systematic study of atmosphere-induced influences}
\authors{Michael Prouza$^{1}$, for the Pierre Auger
Collaboration$^{2}$}
\afiliations{$^{1}$ Nevis Institute and Department of Physics,
Columbia University, New York, N.Y., U.S.A.\\
$^{2}$ Observatorio Pierre Auger, Av. San Mart\'{\i}n Norte 304,
        5613 Malarg\"{u}e, Argentina}
\shortauthors{Pierre Auger Collaboration}
\email{prouza@nevis.columbia.edu}

\abstract{A wide range of atmospheric monitoring
instruments is employed at the Pierre Auger Observatory : two laser 
facilities, elastic lidar stations, aerosol phase
function monitors, a horizontal attenuation monitor, star monitors, weather
stations, and balloon soundings. We describe the impact of analyzed
atmospheric data  
on the accuracy of shower reconstructions, and in particular study 
the effect of the data on the shower energy and the depth
of shower maximum ($X_\text{max}$). These effects have
been studied using the subset of ``golden hybrid'' events --- events observed
with high quality in the fluorescence and surface detector --- used in the
calibration of the surface detector energy spectrum.}

\begin{document}
\maketitle



\section{Introduction}

The Pierre Auger Observatory in Malarg\"{u}e, Argentina is a hybrid facility
that uses its atmospheric fluorescence detector (FD) to obtain calorimetric
estimates of shower energies.  The atmosphere acts both as a calorimeter and a
scintillator. It affects the fluorescence yield of showers and attenuates light
between showers and the FD. The atmosphere over the observatory 
is also dynamic. Therefore, the state of the atmosphere must be
continuously monitored to ensure reliable energy estimates. 

To an excellent approximation, the molecular and aerosol scattering processes
that contribute to the overall attenuation and scattering of light in the
atmosphere can be treated separately.  In Malarg\"{u}e, the molecular component
is determined by regular measurements of several macroscopic
parameters, including altitude profiles of air temperature, pressure, and
density.  It has been shown that daily variations in these parameters have a
small impact on shower energy estimates ($\Delta E/E < 1\%$) and the depth of
shower maximum ($\Delta X_\text{max}\simeq 6\text{ g cm}^{-2}$); hence, the
observations have been incorporated into monthly models for use in the FD
reconstruction~\cite{keilhauer_icrc}.  A more important factor is the effect of
humidity on the fluorescence yield, because the yield provides a
scaling factor for the energy. Preliminary estimates suggest
a $5 - 10$\% effect on the fluorescence yield near the ground, and $<3\%$ at
$4$~km above sea level~\cite{keilhauer_fluor}. This effect
has been incorporated into the reported
uncertainty of the fluorescence yield \cite{spectrum1}.
Aerosols, unlike the molecular component of the atmosphere, are much more
variable, and can change significantly in the course of a few hours.
Therefore, aerosols are systematically measured at all FD sites, and the 
parameters most important for the FD reconstruction are recorded
hourly.  
The aerosol data and the facilities used to collect them are
described in detail elsewhere~\cite{aerosol_icrc}.  In this paper, we discuss
the effect of aerosol measurements on energy and $X_\text{max}$ for an
important subset of observed showers.

\section{Aerosol Measurements}

The presence of aerosols does not influence the air fluorescence yield, so
their primary effect on the shower reconstruction comes from their role in
light attenuation and scattering.  Fluorescence light is emitted isotropically,
so a detector with a field of view $\Delta\Omega$ observing shower
light of intensity $I_0$ will observe a light level

\begin{equation}
  I = I_0 \cdot T_m \cdot T_a \cdot(1+H.O.)\cdot\frac{\Delta\Omega}{4\pi}
\end{equation}

In this expression, $T_m$ is the transmission factor due to molecular
scattering; $T_a$ is the transmission due to aerosol scattering; and $H.O.$ is
a higher-order correction that accounts for the single and multiple scattering
of photons into (or out of) the detector field of view.  Ignoring multiple
scattering effects, $T_a$ and $H.O.$ may be fully characterized using three
independent measurements: the height profile of the vertical aerosol optical
depth (VAOD) $\tau (h)$; the wavelength dependence of the VAOD; and the normalized
aerosol differential scattering cross-section, or phase function, $P(\theta)$.

The VAOD primarily affects light attenuation.  It is defined for each height
above the ground level, so that aerosol transmission $T_{a} (h)$ between the ground 
and height $h$ is 

\begin{equation}
  T_{a} (h) = e^{- \tau(h)} \mbox{.}
\end{equation}

The height profile $\tau (h)$ is measured independently by three elastic backscatter 
lidar stations and 
FD-reconstructed laser tracks provided by the Central Laser Facility (CLF)
\cite{aerosol_icrc,clf_nim,lidars_nim}. With a typical measurement uncertainty
of $\pm0.01$, the CLF measurements are currently used for shower reconstructions.  
However, both the lidars and the CLF 
use monochromatic light sources, so the wavelength
dependence of the VAOD is measured by two other independent instruments: the Horizontal
Attenuation Monitor (HAM) and the robotic astronomical telescope
FRAM~\cite{ham_icrc,fram_icrc}.  This wavelength dependence can be parametrized
in terms of the so-called Angstrom exponent $\gamma$:

\begin{equation}
\tau(\lambda) = \tau_0\cdot
  \left(\frac{\lambda_0}{\lambda}\right)^{\gamma} \text{,}
\end{equation}
where $\tau_0$ is measured at a reference wavelength $\lambda_0=355$~nm.
Observations made between June and December 2006 indicate a typical value
$\gamma=\text{0.7}\pm{0.5}$ for the observatory location.

%

Finally, aerosols not only attenuate light from air showers, but also
scatter Cherenkov light into the FD field of view, contaminating the
fluorescence signal.  The angular distribution of aerosol-scattered
light is given by the aerosol phase function, which we model using 
two free parameters (see \cite{aerosol_icrc}). The parameter $f$ is sensitive 
to the relative strength of forward and backward scattering and $g$, the mean cosine 
of the scattering angle, is the measure of
the asymmetry of scattering. Aerosol Phase Function monitors (APFs) perform
hourly measurements of these aerosol scattering properties at two FD locations
\cite{aerosol_icrc}.
%

\section{Analysis}

The showers of greatest importance to Auger are those measured with high
quality in both the FD and the surface detector (SD), because these are used to
set the energy scale of the overall detector.  For the analysis presented here,
the set of events passing strong quality cuts presented
in~\cite{spectrum1,spectrum2} are used to determine the effect of aerosol
measurements on energy and $X_\text{max}$ estimated by the FD
reconstruction.
Only cloud-free measurements, identified by a strict quality cut on the
FD longitudinal profile, were used in this study.

Several different studies are of interest:

1. The use of aerosols in the reconstruction, compared to the
        use of a pure molecular atmosphere.

2. The propagation of measurement uncertainties in VAOD, $\gamma$, $f$,
        and $g$ in the FD reconstruction, and in particular their effect on
        energy and $X_\text{max}$. 

3. A test of the assumption of atmospheric horizontal uniformity
        used in the FD reconstruction.

  \begin{figure}[b]
  \label{ray_vs_meas}
    \begin{center}
      \includegraphics[width=0.4\textwidth]{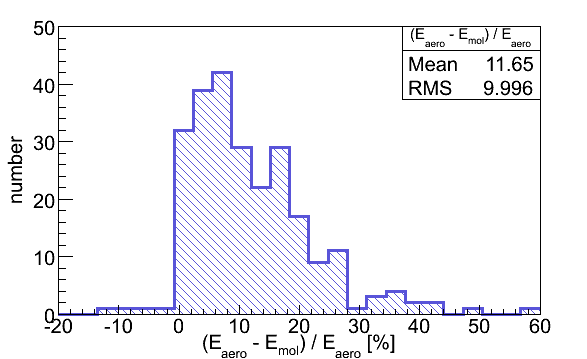}
      \caption{Energy difference for events reconstructed with
measured aerosol parameters (E$_\text{aero}$) and with assumption of no aerosols
(molecular atmosophere; E$_\text{mol}$).}
    \end{center}
  \end{figure}

  \begin{figure*}[th]
  \label{aer_param}
    \begin{center}
      \includegraphics[width=0.32\textwidth]{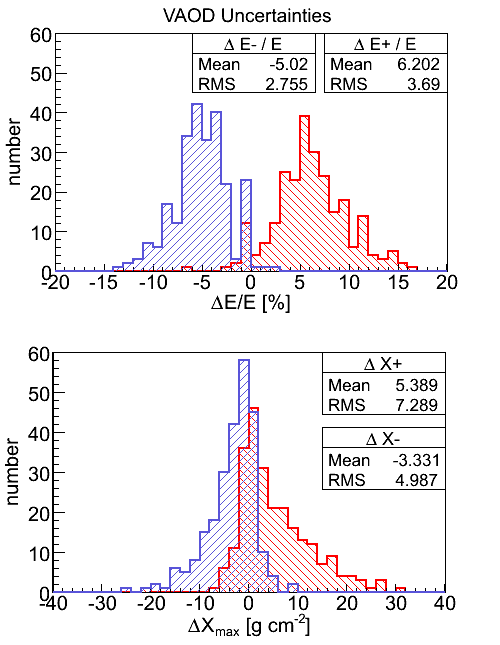}
      \includegraphics[width=0.32\textwidth]{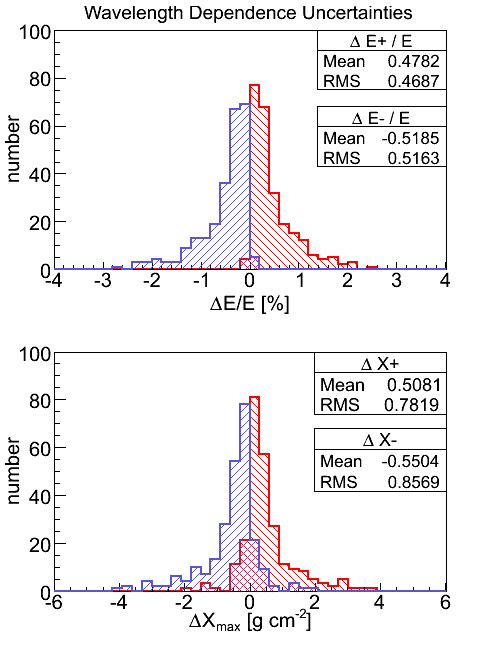}
      \includegraphics[width=0.32\textwidth]{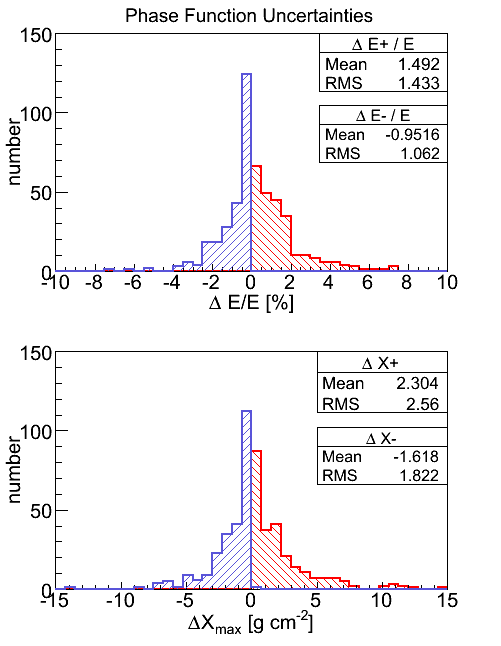}
      \caption{Spread in $E$ and $X_\text{max}$ due to statistical fluctuations
               in aerosol measurements. $\Delta$E$_{+}$ and $\Delta$X$_{+}$ are calculated
using the +1 $\sigma$ uncertainty (on aerosol measurement), $\Delta$E$_{-}$
 and $\Delta$X$_{-}$ using the $-1$ $\sigma$ uncertainty.}
    \end{center}
  \end{figure*}

{\bf (1) Effects of the Presence of Aerosols on Air Shower Detection}

 We have
  compared the reconstruction of showers using real-time aerosol measurements
  with the same events reconstructed using a purely molecular atmosphere.
  As shown in Figure 1, neglecting the presence of
aerosols causes, on average, a 12\% underestimate in
shower energies. Moreover, the long tail in the distribution indicates the
enormous effect of aerosol attenuation on the reconstruction of a
significant fraction of all showers: 15\% of the showers have an
energy correction greater than 
25\%; 6.5\% of events more than 30\%; and 3\% more than 40\%.
Especially for the highest energy events, where the 
statistics are poor, real-time atmospheric calibration is essential. 

Having established the significant influence of aerosols on the reconstruction,
we can also ask if simple parametric models of the aerosol content are
sufficiently accurate.  For example, a two-parameter exponential aerosol density 
profile, characterizing conditions at the site,
 has been considered.  Preliminary studies indicate 
that this parameterization leads to a 4\% overestimate in shower energies (with a 
large spread of 10\%) compared to reconstructions performed with true aerosol 
measurements.

{\bf (2) Uncertainties Introduced by Aerosol Measurements}

We have propagated the measurement uncertainties in the VAOD, phase
function, and wavelength dependence in the hybrid reconstruction.
Figure 2 depicts the contribution of each 
measurement to the
uncertainty in energy and shower maximum.  The VAOD provides
the dominant contribution to the shower uncertainties: 5.5\% for the
statistical uncertainty in energy and 4~g~cm$^{-2}$ for $X_\text{max}$.  
The wavelength dependence and the phase function are significantly less important,
contributing 1\% and 1.3\%, respectively, to the uncertainty of the
energy, and $\sim 2$~g~cm$^{-2}$ to the uncertainty in $X_\text{max}$.

{\bf (3) Evaluation of the Horizontal Uniformity of the Atmosphere}

The atmospheric measurements at the Auger Observatory, while extensive,
are only able to observe conditions at several locations across the 
site.  Therefore, during the reconstruction of events one must assume
that these limited measurements characterize broad regions of the
atmosphere around each FD, or in other words, that the atmosphere
exhibits a large degree of horizontal uniformity.

  \begin{figure}[ht]
  \label{bianca}
    \begin{center}
      \includegraphics[width=0.4\textwidth]{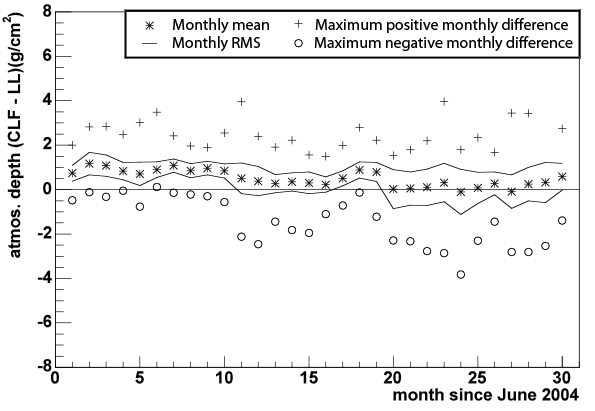}
      \caption{Comparison of monthly atmospheric depth measurements 
at two sites separated by $\sim$30 km.}
    \end{center}
  \end{figure}

The assumption of horizontal uniformity and its effect on the
reconstruction can be tested for both the molecular and aerosol   
components of the atmosphere.  Molecular conditions are observed by
balloon flights and two ground-based weather stations located at Los
Leones and the Central Laser Facility.  
Figure 3 indicates differences in the weather
conditions observed at these sites, and the corresponding effect of  
these differences on shower reconstructions is $< 1\%$ for the shower
energy and $\sim 1$~g~cm$^{-2}$ for $X_\text{max}$.

For the aerosol component of the atmosphere, it is possible to estimate
the effect of horizontal nonuniformities by reconstructing showers using
different sets of aerosol measurements.  For this study, we have  
reconstructed a set of high quality air showers observed by the Coihueco
FD using CLF VAOD profiles measured concurrently at Coihueco and Los
Leones (the distance between these two FD sites is about 40 km). 
As shown in Figure 4, the systematic uncertainty on
shower energies introduced by the assumption of uniformity is $\sim2.5\%$, with
measurement fluctuations contributing $\sim7\%$ to the statistical
uncertainty.  The shower maximum $X_\text{max}$ is also shifted by
$\sim2.5$~g~cm$^{-2}$, with typical statistical uncertainties of $\sim9$~g~cm$^{-2}$.

  \begin{figure}[ht]
  \label{homog}
    \begin{center}
      \includegraphics[width=0.4\textwidth]{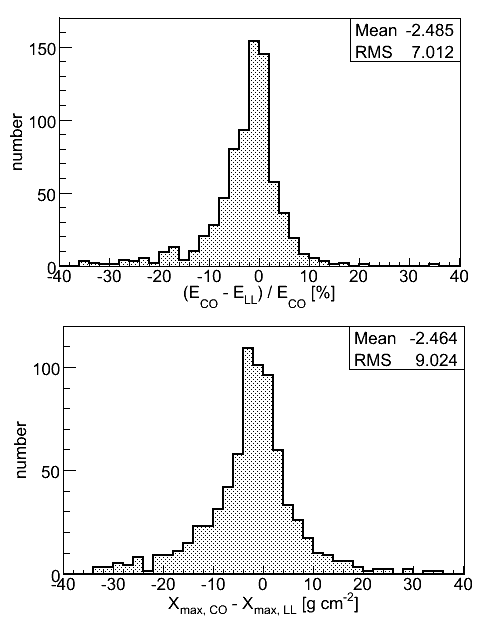}
      \caption{Comparison of energy difference (top) and of shift in
position of shower maximum (bottom) due to the use of non-local
aerosol atmospheric parameters.}
    \end{center}
  \end{figure}

\section{Discussion}

The following table summarizes atmosphere-induced uncertainties 
in the hybrid reconstruction:

\begin{table}[ht]
\begin{center}
{\scriptsize
\begin{tabular}{lrr}
\hline
{\bf Effect} & $\Delta$ E / E & $\Delta$ X$_{\mbox{max}}$\\
\hline
{\bf Molecular:} & \qquad & \qquad \\ 
Horizontal uniformity	& $<$ 1 \% & 1 g cm$^{-2}$ \\ 
Variations in air density profile & 1 \% & 6 g cm$^{-2}$\\
{\bf Aerosols:} & \qquad & \qquad \\
Horizontal uniformity [systematic] & 2.5 \% & 2.5 g cm$^{-2}$\\
Horizontal uniformity [statistical] & 7 \% & 9 g cm$^{-2}$\\
Vertical Aerosol Optical Depth	& 5.5 \% & 4 g cm$^{-2}$ \\
Wavelength Dependence & 1 \% &	1 g cm$^{-2}$ \\
Differential Scattering Cross-section & 1.3 \% & 2 g cm$^{-2}$ \\
\hline
\end{tabular}
}
\end{center}
\caption{Summary of atmosphere-induced uncertainties for
the set of calibration events}
\end{table}

\end{document}